\documentclass[longauth]{aa} % for the long lists of affiliations 
\usepackage{graphicx,color}
\usepackage{natbib}
\bibpunct{(}{)}{;}{a}{}{,} % to follow the A&A style
\usepackage[small]{caption}

\begin{document}

\title{Detection of Very High Energy radiation from HESS~J1908+063 confirms the Milagro unidentified source MGRO~J1908+06}

\subtitle{}

\author{F. Aharonian\inst{1,13}
 \and A.G.~Akhperjanian \inst{2}
 \and G.~Anton \inst{16}
 \and U.~Barres de Almeida \inst{8} \thanks{supported by CAPES Foundation, Ministry of Education of Brazil}
 \and A.R.~Bazer-Bachi \inst{3}
 \and Y.~Becherini \inst{12}
 \and B.~Behera \inst{14}
 %\and M.~Beilicke \inst{4}
 \and W.~Benbow \inst{1}
 %\and D.~Berge \inst{1} \thanks{now at CERN, Geneva, Switzerland}
 \and K.~Bernl\"ohr \inst{1,5}
 \and C.~Boisson \inst{6}
% ...
}

\newpage

\institute{
Max-Planck-Institut f\"ur Kernphysik, P.O. Box 103980, D 69029
Heidelberg, Germany
\and
 Yerevan Physics Institute, 2 Alikhanian Brothers St., 375036 Yerevan,
Armenia
\and
Centre d'Etude Spatiale des Rayonnements, CNRS/UPS, 9 av. du Colonel Roche, BP
4346, F-31029 Toulouse Cedex 4, France
\and
Universit\"at Hamburg, Institut f\"ur Experimentalphysik, Luruper Chaussee
149, D 22761 Hamburg, Germany
\and
Institut f\"ur Physik, Humboldt-Universit\"at zu Berlin, Newtonstr. 15,
D 12489 Berlin, Germany
\and
LUTH, Observatoire de Paris, CNRS, Universit\'e Paris Diderot, 5 Place Jules Janssen, 92190 Meudon, 
France
Obserwatorium Astronomiczne, Uniwersytet Ja
\and
IRFU/DSM/CEA, CE Saclay, F-91191
Gif-sur-Yvette, Cedex, France
\and
University of Durham, Department of Physics, South Road, Durham DH1 3LE,
U.K.
\and
Unit for Space Physics, North-West University, Potchefstroom 2520,
    South Africa
\and
Laboratoire Leprince-Ringuet, Ecole Polytechnique, CNRS/IN2P3,
 F-91128 Palaiseau, France
\and 
Laboratoire d'Annecy-le-Vieux de Physique des Particules, CNRS/IN2P3,
9 Chemin de Bellevue - BP 110 F-74941 Annecy-le-Vieux Cedex, France
\and
Astroparticule et Cosmologie (APC), CNRS, Universite Paris 7 Denis Diderot,
10, rue Alice Domon et Leonie Duquet, F-75205 Paris Cedex 13, France
\thanks{UMR 7164 (CNRS, Universit\'e Paris VII, CEA, Observatoire de Paris)}
\and
Dublin Institute for Advanced Studies, 5 Merrion Square, Dublin 2,
Ireland
\and
Landessternwarte, Universit\"at Heidelberg, K\"onigstuhl, D 69117 Heidelberg, Germany
\and
Laboratoire de Physique Th\'eorique et Astroparticules, CNRS/IN2P3,
Universit\'e Montpellier II, CC 70, Place Eug\`ene Bataillon, F-34095
Montpellier Cedex 5, France
\and
Universit\"at Erlangen-N\"urnberg, Physikalisches Institut, Erwin-Rommel-Str. 1,
D 91058 Erlangen, Germany
% ...
}

\offprints{\\emma@apc.univ-paris7.fr, djannati@apc.univ-paris7.fr}

\date{Received  ; accepted }
 
\abstract{}
{Detection of a $\gamma$-ray source above 300~GeV is reported,
confirming the unidentified source MGRO~J1908+06, discovered by
the Milagro collaboration at a median energy of 20~TeV.} {The source
was observed during 27~h as part of the extension of the
H.E.S.S. Galactic plane survey to longitudes $>$\,30$\degr$. } 
{HESS~J1908+063 is detected at a significance level of 10.9$\sigma$
with an integral flux above 1~TeV of (3.76$\pm$0.29$_{\rm~stat}\pm$
0.75$_{\rm sys})\times$10$^{-12}$ph~cm$^{-2}$s$^{-1}$, and a spectral
photon index $\Gamma$ = 2.10$\pm$0.07$_{\rm~stat}\pm$ 0.2$_{\rm
sys}$. The positions and fluxes of HESS~J1908+063 and MGRO~J1908+06
are in good agreement. Possible counterparts at other wavelengths and
the origin of the $\gamma$-ray emission are discussed. The nearby
unidentified GeV source, GRO~J1908+0556 (GeV) which also remains
unidentified and the new Fermi pulsar 0FGL~J1907.5+0617, may be
connected to the TeV source.}  {}

\authorrunning{Aharonian et al.}
\titlerunning{Detection of HESS~J1908+063 confirms MGRO~J1908+06}

\maketitle

%%%%%%%%%%%%%%%%%%%%%%%%%%%%%%%%%%%%%%%%%%%%%%%%%%%%%%%%%%%%%%%%%%%%%%%%%%%%%%%%%
\section{Introduction}
%%%%%%%%%%%%%%%%%%%%%%%%%%%%%%%%%%%%%%%%%%%%%%%%%%%%%%%%%%%%%%%%%%%%%%%%%%%%%%%%%

Very High Energy (VHE) $\gamma$-rays probe sites of particle
acceleration to ultra-relativistic energies. Observations with
H.E.S.S. (High Energy Stereoscopic System) in the galactic domain,
and, in particular, the Galactic plane survey (GPS) of the central
region of the Milky Way (l $\simeq-30{\degr}$ to 30${\degr}$, b
$\simeq -3^{\circ}$ to 3$^{\circ}$, 2004$-$2005), and its extension to
longitudes l $\simeq 30{\degr}$ to 60${\degr}$ and l $\simeq
280{\degr}$ to 330${\degr}$ during 2005$-$2007, have resulted in the
discovery of more than 40 sources (\citet{HESSScan, HESSScanII,
Unidentified}).
%Firmly established classes of galactic $\gamma$-ray sources 
A large majority of these are extended sources which renders their
identification difficult, except when they exhibit a clear morphology
or correlation with an identified object at other wavelengths.  In
addition to firmly identified classes, i.e. shell-type supernova
remnants (SNRs), pulsar wind nebulae (PWNe) and X-ray binary systems
(XRBs), strong evidence has been obtained for interaction of enhanced
fluxes of cosmic rays with target material in the surroundings of SNRs
(\cite{HESSW28, HESSCTB37a, MagicIC443}), as well as in the central
100 pc of the Galaxy (\cite{GCDiffuse}).  Nevertheless the majority of
galactic $\gamma$-ray sources discovered by H.E.S.S. still lack firm
identification (\cite{Unidentified}).
%(\cite{HESSSurveyICRC07}).

The Milagro collaboration has recently published its sky survey
results after seven years of operation (2358 days of data observed
from a site at $35{\degr}$ N latitude, \cite{MILAGRO}), announcing the
discovery of three low-latitude sources and 4 lower significance
hot-spots. One of the sources, MGRO~J1908+06, is detected at
8.3$\sigma$ (pre-trial) confidence level, with a differential flux of
(8.8$\pm$2.4$_{\rm~stat}\pm$2.6$_{\rm~sys})\times10^{-15}$~TeV$^{-1}$cm$^{-1}$s$^{-1}$,
derived at $\sim$ 20~TeV assuming a spectral photon index of 2.3.  Its
angular extension remains unknown, but is bounded to a maximum
diameter of 2.6${\degr}$.  MGRO~J1908+06 is located near 40$\degr$
longitude
(l=$40\degr$24$\arcmin$$\pm$6$\farcm0_{\rm~stat}\pm$18$\farcm0_{\rm
sys}$,
b=$-1\degr$0$\arcmin$$\pm$6$\farcm0_{\rm~stat}\pm$18$\farcm0_{\rm
sys}$), which is covered by the extended H.E.S.S. GPS.

The detection of an extended $\gamma$-ray source, HESS~J1908+063, in a
compatible direction with that of MGRO~J1908+06, is reported here. The
lower energy threshold of H.E.S.S. and its superior angular resolution
are used advantageously to derive the position, morphology and the
spectrum of the source from 0.3 up to 30~TeV. After the description of
the observations, data analysis and results in
section~\ref{OBSDETECT}, the H.E.S.S. source characteristics, given in
section~\ref{MORPHSPECTRUM}, will be compared to that of
MGRO~J1908+06. Its possible counterparts shall finally be discussed in
section \ref{DISCUSS}.

\begin{figure}[!t]% [!tp]
\begin{center}
\includegraphics[width=0.48\textwidth,angle=0,clip]{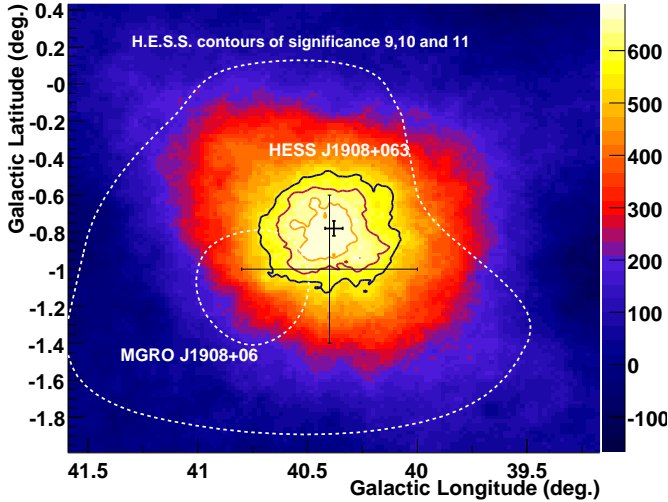}
\end{center}
\caption{Gaussian-smoothed ($\sigma=0.5{\degr}$) excess map of the 2.4${\degr}\times$2.4${\degr}$ field of view around the position of
  HESS~J1908+063. The colored contours show the pre-trial significance
  levels for 9, 10 and 11$\sigma$.  The dotted white lines show the
  Milagro significance contours for 8$\sigma$ (inner) and 5$\sigma$
  (outer contour). The black crosses mark the best H.E.S.S. and
  Milagro fitted positions, respectively, the error bars being the
  linear sum of the statistical and systematic errors. }
%white circle shows the $0.5^{\circ}$ integration radius used for the
%spectrum derivation. 

\label{fig1}
\end{figure}
%

%%%%%%%%%%%%%%%%%%%%%%%%%%%%%%%%%%%%%%%%%%%%%%%%%%%%%%%%%%%%%%%%%%%%%%%%%%%%%%%%%
%\section{Observations, Analysis \& Results}
\section{Observations and detection of HESS~J1908+063}
\label{OBSDETECT}
%%%%%%%%%%%%%%%%%%%%%%%%%%%%%%%%%%%%%%%%%%%%%%%%%%%%%%%%%%%%%%%%%%%%%%%%%%%%%%%%%

H.E.S.S. is an array of four Imaging Atmospheric Cherenkov telescopes
and is located in the Khomas Highland in Namibia. Each telescope has a
mirror area of 107~m$^2$ (\cite{mirrors}) and a total field of view of
5$\degr$ (\cite{camera}), well suited for the study of extended
sources.  The system works in a coincidence mode
(e.g. \cite{trigger}), requiring at least two of the four telescopes
to have triggered. Its angular resolution reaches $\sim$~5' per event
and its sensitivity for a point-like source is
$~2.0\times10^{-13}$~ph~cm$^{-2}$s$^{-1}$ (1$\%$ of the Crab Nebula
flux above 1~TeV) for a 5$\sigma$ detection within 25~hours
observation time.

Observations near Galactic longitude l=40$\degr$ were first performed
during June 2005 and then from May to September 2006, as part of the
extension of the H.E.S.S. GPS to l $\simeq$ 30${\degr}$ to
$60{\degr}$. Subsequent to the first detection of HESS~J1908+063,
follow-up observations were made in wobble mode during 2007, where the
source direction was positioned $\pm$0.5$^o$ in declination relative
to the center of the field of view of the camera. Due to the large
field of view and uniform response of the H.E.S.S. cameras, the use of
Wobble mode allows for both on-source observations and simultaneous
estimation of the background induced by charged cosmic rays, since the
background can be estimated from different regions in the same field
of view. The direction of the offset was alternated in successive
scans to reduce systematic effects. Due to the survey-mode
observations the source is offset from the field of view center at
different angular distances. Observations with an offset of more than
2.0$\degr$ (2.5${\degr}$) were not used for the determination of the
spectrum (for the sky maps). The total dead-time corrected, quality
and offset selected (2.5${\degr}$) data set amounts to 27~hours and
has an average offset of 1.0${\degr}$. The zenith angle ranges from 30
to 55${\degr}$, leading to a mean energy threshold of $\sim$300~GeV,
for a cut on the image size of 80 photoelectrons (p.e.), and
$\sim$600~GeV for a tighter cut of 200 p.e.

After calibration, the standard H.E.S.S. event reconstruction scheme
was applied to the data (\cite{HESSCrab}). In order to reject the
background of cosmic-ray showers, $\gamma$-ray like events were
selected using cuts on image shape parameters scaled with their
expected values obtained from Monte Carlo simulations. As described in
\cite{HESSScanII}, two different sets of cuts were applied. Cuts optimized for a hard photon spectrum and a
weak source with a rather tight cut on the image size of 200 p.e.,
which achieve a maximum signal-to-noise ratio, were applied to study
the morphology of the source, while for the spectral analysis, the
image size cut was loosened to 80 p.e. in order to cover the maximum
energy range. The background estimation (described in \cite{HESSBack})
for each position in the two-dimensional sky-map was computed from a
ring with an ({\em a-priori}) increased radius of $1.0{\degr}$, as
compared to the standard radius of 0.7$\degr$, in order to account
with the large source diameter. The width of the ring was selected to
reach a background area four times that of the on source area.
%Also events coming from known nearby
%sources were excluded to avoid contamination of the background.

Fig.~\ref{fig1} shows the Gaussian-smoothed ($\sigma$=0.5$\degr$ to
compare to the Milagro point spread function, PSF) excess map for a
size cut on the images above 200 p.e. An excess of 689 events at a
pre-trial significance of 12.1 $\sigma$ is obtained using an
integration radius of 0.5$\degr$. A conservative estimate of the
trials, following the same procedure as that described in
\cite{HESSScanII} but taking into account the larger area
covered by the extended GPS, yields a post-trial significance of 10.9
$\sigma$ (for 525000 trials).
 
To evaluate the extension and the position of the source, the
uncorrelated excess sky-map was fitted to a symmetrical
two-dimensional Gaussian function, convolved with the H.E.S.S. PSF
(PSF$\sim$0.08$^{\rm o}$ above 200 p.e.). The best-fit position lies
at l=40$\degr$23$\arcmin$9$\farcs$2$\pm$2$\farcm4_{\rm~stat}$ and
b=-0$\degr$47$\arcmin$10$\farcs$1$\pm$2$\farcm4_{\rm~stat}$, with a
systematic error of 20'' per axis (\cite{AngularRes}), and the
intrinsic extension derived is
$\sigma_{src}=0.34{\degr}^{+0.04}_{-0.03}$.

\begin{figure}[!t]
\begin{center}
\includegraphics*[width=0.45\textwidth,angle=0,clip]{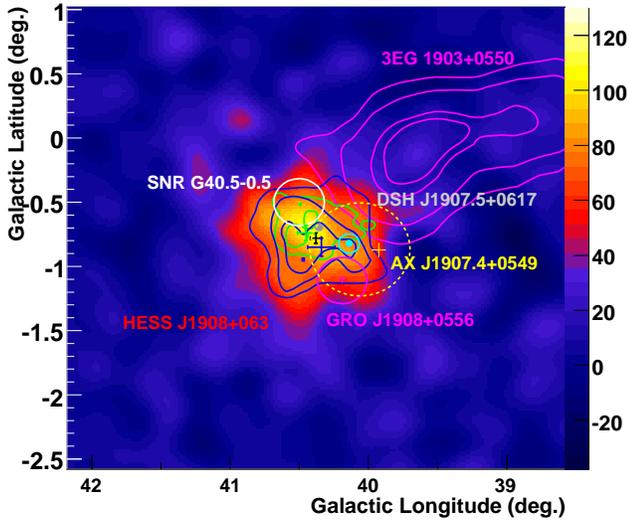}
\end{center}
\caption{Gaussian-smoothed ($\sigma=0.11{\rm \degr}$) excess map of the 3.6${\degr}\times3.6{\degr}$ field of view around the position of HESS~J1908+063. The H.E.S.S. significance contours (for an integration radius of 0.22$\rm \degr$) above
4$\sigma$ are superimposed and marked in green (for the energy range
between 0.7 to 2.5 TeV) and blue (for energies above 2.5 TeV)
contours. The green and blue crosses show the two best fitted
positions for the two excesses at the two different energy ranges, the
bars being the statistical errors. The possible counterparts are also
shown (see section \ref{DISCUSSCounterparts}).}
\label{fig2}
\end{figure}

%%%%%%%%%%%%%%%%%%%%%%%%%%%%%%%%%%%%%%%%%%%%%%%%%%%%%%%%%%%%%%%%%%%%%%%%%%%%%%%%%
%\section{Observations, Analysis \& Results}
\section{Morphology and spectrum}
\label{MORPHSPECTRUM}
%%%%%%%%%%%%%%%%%%%%%%%%%%%%%%%%%%%%%%%%%%%%%%%%%%%%%%%%%%%%%%%%%%%%%%%%%%%%%%%%%

To study the morphology of the source, a cut of 200~p.e. on the shower
size was applied to optimize the signal-to-background ratio and
angular resolution, and the resulting sky map was Gaussian-smoothed
($\sigma$=0.11$^{\rm o}$). As can be seen in Fig.~\ref{fig2}, the
excess event map departs apparently from a Gaussian shape, but the
significance of this effect is marginal. When excess events are
separated in two different energy bands, 0.7 to 2.5~TeV, and that
above 2.5~TeV (the two bands were selected ({\em a-priori}) such that
the signal-to-noise ratio remains constant under the hypothesis of a
source with a hard spectrum photon index of 2.2), the emission peaks
of the corresponding excess maps show a slight offset (0.16$\degr$
apart) with respect to each other, following roughly the morphology of
the total excess map.
Two circular regions of radius 0.2$\degr$ were selected around the
positions of the low and high energy fitted centroids. The energy
spectrum was derived in these two regions (see below for details of
the procedure). The results of these spectral analysis show the
expected tendency, i.e. the spectral index derived at the high energy
peak is harder than the one at the low energy region, with a
difference of 0.17. Nevertheless the systematic error of the
measurements ($\pm$0.2) forbids any conclusion on the separation in
two sources/peaks.

\begin{figure}[!t] 
\begin{center}
\includegraphics[width=0.5\textwidth]{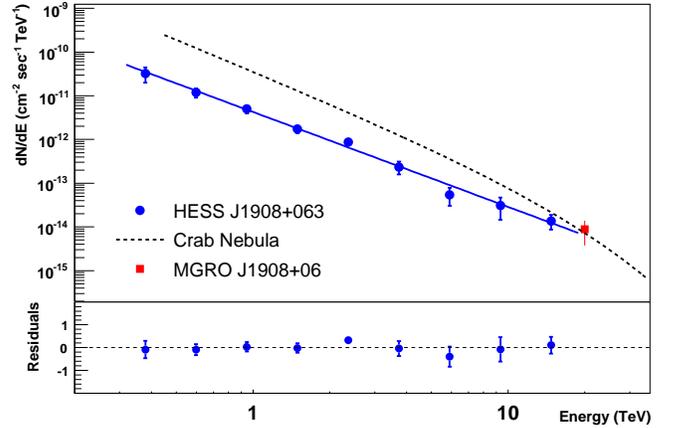}
\end{center}
\caption{Differential energy spectrum of HESS~J1908+063 measured in
  the energy range 300~GeV-30 TeV. The differential flux of
  MGRO~J1908+06 at 20~TeV is shown in red. The black dotted line
  represents the Crab Nebula energy spectrum measured by H.E.S.S. The
  residuals to the power-law fit are shown in the lower panel.}
\label{spectrum} 
\end{figure}

\begin{figure}[!t]
\begin{center}
\includegraphics[width=0.55\textwidth,angle=0,clip]{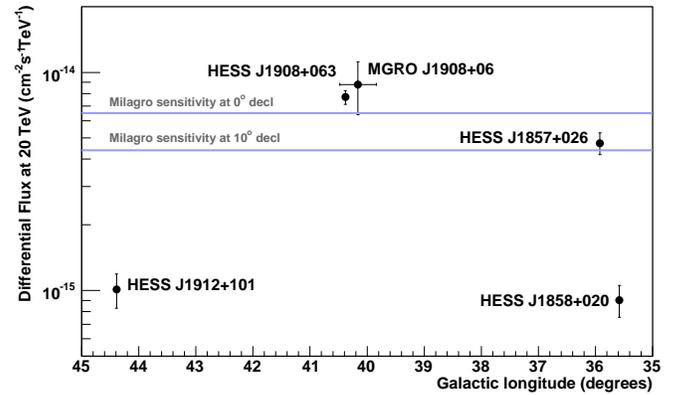}
\end{center}
\caption{Differential flux at 20~TeV vs. galactic longitude for
  sources detected by H.E.S.S. and Milagro for $l \in[35{\degr}, 45{\degr}]$. The
  blue lines show the differential sensitivity of Milagro for 0${\degr}$ and
  10$^o$ declination, which correspond to the declination range of 
  Milagro observations for these galactic longitude 
  values.  
}
\label{comparison}
\end{figure}

%Therefore 
The energy spectrum was derived considering a single source within an
integration radius of $0.5{\degr}$ (to take into account the angular
extension of the source), centered on the best-fit position, by means
of a forward-folding maximum likelihood fit (\cite{CATSpectrum}). The
background was evaluated from positions in the field of view with the
same radius and same offset from the pointing direction as the source
region. The spectrum, shown in Fig.~\ref{spectrum}, is well fitted
with a simple power-law function with a hard photon index of
$2.10\pm0.07_{\rm~stat}\pm 0.2_{\rm sys}$ and a differential flux at
1~TeV of ($4.14 \pm 0.32_{\rm~stat} \pm 0.83_{\rm sys})\times
10^{-12}$ TeV$^{-1}$cm$^{-2}$~s$^{-1}$. The syste matic error on the
flux is conservatively estimated from simulated data to be 20$\%$
while the photon index has a typical systematic error of $\pm$0.2. The
integrated flux above 1~TeV corresponds to 17$\%$ of the Crab Nebula
flux above that energy (\cite{HESSCrab}).

An exponential energy cut-off has been searched for in the data. The
flux was fit by a power law with an exponential cutoff of the form
F(E)=N$\rm _o$(E/1TeV)$^{-\rm \Gamma}$exp(E/E$_{\rm cut}$), where
N$_o$ is the flux normalization and $\Gamma$ the photon index. There
is no indication for such a cut-off and a lower limit of $E_{\rm
cut}>$19.1~TeV ($\pm$15$\%$ systematic error) can be derived at 90$\%$
confidence level.

%%%%%%%%%%%%%%%%%%%%%%%%%%%%%%%%%%%%%%%%%%%%%%%%%%%%%%%%%%%%%%%%%%%%%%%%%%%%%%%%%
\section{Discussion}
\label{DISCUSS}
%%%%%%%%%%%%%%%%%%%%%%%%%%%%%%%%%%%%%%%%%%%%%%%%%%%%%%%%%%%%%%%%%%%%%%%%%%%%%%%%%

%HESS=MGRO
\subsection{Comparison with MGRO~J1908+06}
The fitted positions of HESS~J1908+063 and MGRO~J1908+06 and their
significance contours are in good agreement as shown in
Fig.~\ref{fig1}. The association is supported by a comparison of the
H.E.S.S. derived spectrum to the Milagro differential flux at 20~TeV
(see Fig.~\ref{spectrum}), where the Milagro point error bar is the
linear sum of the statistical and systematic errors while the
H.E.S.S. points error bars represent only the statistical error. The
latter was obtained using a larger integration radius of 1.3${\degr}$
(90$\%$ CL), compared to 0.5$\degr$ for HESS J1908+063. The positional
and flux agreements (the systematic errors on absolute flux
measurements are about 20-30$\%$ for the two instruments) are
consistent with the fact that the emission seen in the two instruments
comes from the same source, and that no other significant emission
from outside the H.E.S.S. integration radius of 0.5$\degr$ contributes
to the flux of MGRO J1908+06.

%Other hess sources: extension of the spectrum 

\subsection{Milagro Detectability - Neighbouring H.E.S.S. sources}

Since the detection of HESS~J1908+063 confirms for the first time with
a Cherenkov telescope one of the Milagro sources, the case of other
H.E.S.S. sources at galactic longitudes to the North of 30${\degr}$,
which therefore fall into the overlap region with the Milagro sky
survey, is relevant in this context. These are HESS~J1857+026,
formerly unidentified but for which a potential counterpart,
PSR\,J1856+0245, has been recently discovered (\cite{Unidentified,
Hessels08}) the still unidentified HESS~J1858+020
(\cite{Unidentified}), and the PWN candidate HESS~J1912+101
(\cite{HESSJ1912}).

In order to investigate if these sources were susceptible to be
detected by Milagro, their differential flux, extrapolated to 20~TeV,
is compared to the Milagro sensitivity at ecliptic declinations of
0$^{\circ}$ and 10$^{\circ}$ (taken from \cite{MILAGRO}) in
Fig.~\ref{comparison}.  HESS~J1857+026, with a much smaller angular
extension ($\sim\,0.11\degr$), a higher differential flux at 1~TeV
($6.1 \times 10^{-12}$ TeV$^{-1}$cm$^{-2}$~s$^{-1}$) but a slightly
softer spectrum ($2.39\pm0.08_{\rm~stat}$), seems at the limit of
detectability, while the two others appear to be well below the
Milagro sensitivity. The spectral index of HESS~J1858+020 is hard
($2.17\pm0.12_{\rm~stat}$), but its differential flux is an order of
magnitude below the other sources. HESS~J1912+101, is closer to
HESS~J1908+063 in angular extension (0.26$\degr$) with a comparable
differential flux at 1~TeV ($\sim 3.5 \times 10^{-12}$
TeV$^{-1}$cm$^{-2}$~s$^{-1}$), but has a much steeper spectrum index
($2.7\pm0.2_{\rm~stat}$). This explains the dramatic drop in its
differential flux at 20~TeV as compared to HESS~J1908+063.

\begin{figure}[!t]% [!tp]
\begin{center}
\includegraphics[width=0.45\textwidth,angle=0,clip]{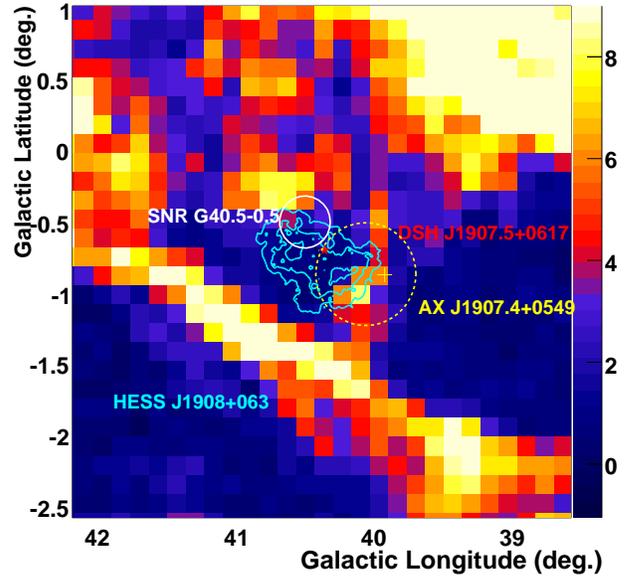}
\end{center}
\caption{On the color scale background, the sky-map in $^{12}$CO(J = 1-0) integrated in the velocity range between 25.3 and 30.5 km/s around the position of the H.E.S.S. source, in blue contours. Possible counterparts are also shown (see section \ref{DISCUSSCounterparts}).}
\label{fig3}
\end{figure}

\subsection{Search for Counterparts}
\label{DISCUSSCounterparts}

Potential counterparts for HESS~J1908+063 are shown in
Fig.~\ref{fig2}: the Galactic open cluster candidate DSH J1907.5+0617,
marked with a grey point; the SNR~G40.5$-$0.5, whose angular extension
is marked with a white circle; the EGRET source 3EG~1903+0550, in
purple contours corresponding to 99, 95, 68 and 50$\%$ confidence
levels and considered as possibly associated with the SNR
(\cite{Hartman99}); the ASCA source AX~J1907.4+0549, shown in light
yellow (the dotted circle shows its observation field of view), was
discovered as a result of the systematic search for identification of
EGRET sources (\cite{Roberts01}) and was proposed as a plausible
counterpart to the other yet unidentified EGRET source in the field,
GRO~J1908+0556 (GeV) (\cite{olaf97,lamb97}); the latter is marked with
a cross in purple together with its 1$\sigma$ position error. Recently
a new pulsar, 0FGL~J1907.5+0602, detected by the Fermi-LAT instrument
(\cite{fermi}) was reported (cyan circle) close to both the TeV and
EGRET sources.

The 26$\arcmin$ shell SNR~G40.5$-$0.5 is in spatial coincidence with
the northeastern part of the H.E.S.S. source. \cite{Downes80}, using
various $\Sigma-$D relations derived a linear diameter of 40--65 pc,
an age of (2--4)$\times 10^4$ yr, and estimated the distance to the SNR
to be in the range 5.5 to 8.5 kpc, corresponding to a location either
in the inter-arm region between the Scutum and the Sagittarius arms,
or on the inner edge of the latter.
\cite{Yang06} investigated the distribution of molecular gas
around the SNR direction through the $^{12}$CO (J = 1--0) transition
line, and provided evidence for interaction between the SNR and its
neighboring dense ISM at a central velocity $\rm V_{\rm LSR}$=55 km/s.
The corresponding kinematic distance of 3--3.4~kpc implied a source
diameter of 25 pc and a younger age, at variance with the initial
estimates of \cite{Downes80}.

The association of the H.E.S.S. source to this SNR is not
straightforward due to the fact that the angular size ($>$ 40$\arcmin$
FWHM) of the VHE source is significantly larger than the 26$\arcmin$
size of the shell. A scenario consisting of a SNR--molecular cloud
association, such as that possibly at work for SNR~W~28
(\cite{HESSW28}) where the $\gamma$-rays are produced through
interactions of accelerated cosmic rays with molecular matter in the
vicinity of the source, along with the contribution of a nearby
unresolved source to the south-west could lead to such a difference in
angular extension. However, the dense region reported by \cite{Yang06}
is located north of the SNR and does not correspond to the observed
$\gamma$-ray emission.

The study of the $^{12}$CO (J = 1--0) data (\cite{Dame}) in the line
of sight of HESS~J1908+063 confirms this discrepancy. However, in the
velocity range $\rm V_{\rm LSR}$=25--30~km/s, corresponding to a much
closer distance of 1.5--1.8 kpc --that is the closer part of the
Sagittarius arm-- an interesting feature might be present: the VHE
source is aligned with a very low density region resembling a bubble
left by a supernova explosion (Fig.~\ref{fig3}) or stellar
winds. \cite{Kronberger06} reported a candidate for a Galactic open
cluster, DSH J1907.5+0617, whose position is compatible with the
H.E.S.S. source best fitted one. Deeper studies of this candidate are
necessary to further clarify its nature and establish a possible
connection with the VHE source and $^{12}$CO (J = 1--0) data.
%The white circle marks the angular extension and position of SNR G040.5-00.5 while in yellow is marked the position of the ASCA source AX J1908+0556 and its field of view, partially covering the H.E.S.S. observations.

The 2--10~keV ASCA GIS image, which led to the discovery of
AX~J1907.4+0549, partially covers the field of view of the
H.E.S.S. source. Another X-ray observation with Swift/XRT was recently
reported (\cite{ATEL1251}), following the announcement of a faint
radio source in archival data (\cite{ATEL1247}) near the
HESS~J1908+063 centroid (RA=19h08m03.73s,
DEC=+06$\degr$18$\arcmin$22$\farcs$4), but resulted only in a
3$\sigma$ upper-limit for a point-like source on the unabsorbed
0.3--10 keV flux of $2.8\times 10^{-13}$ erg cm$^{-2}$
s$^{-1}$. AX~J1907.4+0549 consists of two unresolved peaks, possibly
due to a diffuse emission (\cite{Roberts01}), and lies at the western
edge of HESS~J1908+063, 0.33$\degr$ to its fitted centroid.  If the
X-ray source is indeed extended, it could be a PWN candidate, although
up to now no pulsar has been found in this direction
(\cite{Roberts02}). Assuming there is a PWN in X-rays, its association
to the VHE source would imply an angular offset of 0.48$\degr$. This
type of morphology is not unusual: other PWNe associations such as
HESS~J1825-137, MSH~15-52, HESS~J1718-385, etc. (see
e.g. \cite{Lemiereicrc07}) exhibit such a configuration which is
explained either by the expansion of the SNR in an inhomogeneous
medium and/or the proper motion of the pulsar (see
e.g. \cite{blondin01}). Nevertheless the ASCA X-ray data statistics is too
scarce to establish the existence of a PWN on its own.
% and the hypothesis of an
%offset nebula association is for the moment very speculative.  

On the other hand, the recently discovered Fermi-LAT pulsar,
0FGL~J1907.5+0602, lies at an offset of 0.26$\degr$ relative to the
H.E.S.S. source fitted centroid and is likely to be associated with
it; if so, this would also imply an asymmetrical PWN morphology.
%although it has not been found yet in other wavelengths. 
Previously, in the same energy range, two unidentified EGRET objects
were found in the vicinity of HESS~J1908+063. The harder source, GRO
J1908+0556 (GeV), or alternatively GeV J1907+0557 (the positions of
the two sources are compatible within errors), is consistent with the
position of the Fermi-LAT source and lies within a compatible distance
to the centroid of the H.E.S.S. source, roughly two times the EGRET
68\% position measurement error. If the TeV source and the Fermi
pulsar are indeed related, a sharp cutoff in the photon spectrum of
the latter is expected and no contribution at TeV energies is
expected. The integral flux $>$1~GeV reported for GRO J1908+0556 (GeV)
($6.33\times10^{-8}~{\rm ph~cm}^{-2}{\rm s}^{-1}$) is almost twice
that of the Fermi source ($3.74\times10^{-8}~{\rm ph~cm}^{-2}{\rm
s}^{-1}$). If the GRO and Fermi sources are associated together, the
larger flux of the former could be due to unpulsed PWN emission,
although EGRET calibration issues at several GeV, outlined recently
(\cite{FermiVelaPsr}), may have lead to an overestimation of its flux.
Also, a naive extrapolation of the H.E.S.S. spectrum to lower energies
($>$1~GeV) leads to a lower flux ($7.5\times10^{-9}~{\rm
ph~cm}^{-2}{\rm s}^{-1}$) than the flux difference between the EGRET
source and the Fermi pulsar. Nonetheless the association of the VHE
source to either the EGRET GeV source and/or the Fermi pulsar remains
likely.
% although a coincidence by chance is also
%plausible (\cite{FunkGeVTeVConnection08}). 
%A detailed study of the new
%Fermi-LAT pulsar in comparison with the TeV source will shed light on
%the likely association of these sources, and therefore the possible
%PWN powering the gamma-ray source.

\subsection{Origin of the VHE emission}

TeV $\gamma$-ray emission manifests the presence of ultra-relativistic
particles and could be produced either through Inverse Compton (IC)
scattering of the CMBR, IR and/or star-light seed photons by
electrons, or from the decay of neutral pions resulting from
proton-proton (and other nuclei) interactions. The fact that the
spectrum of HESS~J1908+063 extends up to energies of at least $\mathrm
E_{\gamma}~=$~19~TeV, supported by the Milagro detection, implies
the presence of either electrons of energy $\mathrm E_e \simeq
18\,\times\mathrm \mathrm E_{\gamma}^{1/2} \simeq 80$ TeV (if
neglecting the $\sim$30$\%$ drop in the cross section due to the
Klein-Nishina effect), or protons up to $\mathrm E_p
\simeq \mathrm 10\,\times\mathrm \mathrm E_{\gamma}\sim$ 200 TeV (for typical proton spectra,
%on average roughly 10\% of the parent particle energy goes into production of $\gamma$-rays, 
see e.g. \cite{Kelner}). The available multi-wavelength data do not
allow to distinguish the nature of the particles at the origin of the
$\gamma$-ray emission yet. The X-ray upper-limit of $2.8 \times
10^{-13}$ erg cm$^{-2}$ s$^{-1}$ (\cite{ATEL1251}) is calculated for a
point-like source, and given the large size of the VHE source, it does
not allow to constrain the magnetic field strength for a leptonic
scenario. If one assumes the association with the Fermi pulsar and an
IC emission by electrons, the extension of the spectrum up to 19~TeV
would imply a PWN age of ~5-10 kyr for a magnetic field in the 5 to 3
$\mu$G range (see e.g., eq (6) in \cite{deJagerDjannati08}). On the
other hand, given that the extrapolation of the H.E.S.S. source
spectrum to lower energies is well below the EGRET source flux, a
continuity of the spectral energy distribution from GeV to tens of TeV
remains possible, in which case a hadronic origin of the $\gamma$-ray
emission would constitute also a plausible scenario
(\cite{FunkGeVTeVConnection08}).

%
%______________________________________________________________

\section{Conclusions}

A new H.E.S.S. source, HESS~J1908+063, has been detected above 300 GeV
at a post-trial significance of 10.9$\sigma$ during the extended HESS
Galactic plane survey. It is rather a bright source, with a flux of
17$\%$ of the Crab Nebula. It has a large angular size of
$\sigma$=0.34${\degr}$ and shows a hard spectrum with a photon index
of 2.10$\pm$0.07$_{\rm stat}\pm$0.2$_{\rm sys}$.  The centroid
position and the flux of HESS~J1908+063 are compatible with those of
the unidentified source, MGRO~J1908+06, reported by the Milagro
collaboration at the median energy of 20~TeV. The two sources can
hence be considered as identical. For the first time, one of the
sources discovered by Milagro is confirmed by an Imaging Atmospheric
Cherenkov telescope. The comparison of the fluxes of H.E.S.S. sources
which are covered by the Milagro sky survey to the latter's
sensitivity shows that only another source, HESS~J1857+026, is at the
Milagro detection limit. The hard spectrum of HESS J1908+063 combined
with the detection by Milagro of MGRO J1908+06 at median energies of
20 TeV imply the presence of either electrons, or protons, up to ~80
or ~200 TeV, respectively.

The association with SNR~G40.5--0.5 to the north-east is not excluded
but the larger angular extension of the TeV emission should then find
an explanation in terms of either the contribution of unresolved
sources or the interactions of ultra-relativistic particles with
molecular matter in the vicinity of the SNR. The open cluster
candidate DSH J1907.4+0549 is positionally compatible with the
H.E.S.S. source position, and could be possibly associated to a
``void'' in the $^{12}$CO (J = 1--0) data. If better measurements
support the association to the open cluster, another step could be
made towards establishing this class of objects as VHE emitters
(\cite{Westerlund2}).

On the other hand, the recently reported $\gamma$-ray pulsar,
0FGL~J1907.5+0602, lies close to HESS~J1908+063 centroid
(0.26$\degr$), and could well be associated to the latter in an
offset-type PWN scenario, if the system age is of order of 10000
years. The unidentified GeV source GRO~J1908+0556 (GeV) lies also at a
compatible position to those of the TeV source and the Fermi pulsar
and hence could be related to both. The measurement of the unpulsed
GeV component as well as further multiwavelength observations are
critical for investigating such association and the nature of
HESS~J1908+063/MGRO~J1908+06.

\begin{acknowledgements}
The support of the Namibian authorities and of the University of Namibia
in facilitating the construction and operation of H.E.S.S. is gratefully
acknowledged, as is the support by the German Ministry for Education and
Research (BMBF), the Max Planck Society, the French Ministry for Research,
the CNRS-IN2P3 and the Astroparticle Interdisciplinary Programme of the
CNRS, the U.K. Science and Technology Facilities Council (STFC),
the IPNP of the Charles University, the Polish Ministry of Science and 
Higher Education, the South African Department of
Science and Technology and National Research Foundation, and by the
University of Namibia. We appreciate the excellent work of the technical
support staff in Berlin, Durham, Hamburg, Heidelberg, Palaiseau, Paris,
Saclay, and in Namibia in the construction and operation of the
equipment.
\end{acknowledgements}

\bibliography{J1908p063_V10}

\end{document}